# Interfacial optical absorptance of air-water interfaces

Preston Bohm[†], Mingjun Li[†], Akanksha K. Menon[*], and Zhuomin M. Zhang[*]

*George W. Woodruff School of Mechanical Engineering, Georgia Institute of Technology, Atlanta, Georgia 30332, USA*

[†] Both authors contributed equally to this work
[*] Corresponding author: akanksha.menon@me.gatech.edu, zhuomin.zhang@me.gatech.edu

The photomolecular effect has been hypothesized to enhance evaporation of water at visible wavelengths. This study develops a new measurement technique to investigate the presence and magnitude of the photomolecular effect at the liquid-vapor interface of water. The experiment aims to detect surface absorptance through the comparison of polarized reflectance for substrates with and without a layer of water on top. The ratio of this polarized reflectance is then used as an indicator of interfacial absorptance or attenuation. Lightly doped silicon was used as a dielectric substrate based on its stability and use as a common reference. Platinum was used as a metal substrate for its high reflectance and frequency invariant optical properties. The pseudo-optical properties of both substrates were measured using ellipsometry. The experimental results are compared to a multilayer reflectance model to determine the theoretical sensitivity of the reflectance ratio, thereby guiding angle selection for the experiments. The measurements demonstrated close agreement with a classical model that assumes zero surface absorptance. The classical model is then extended to include surface absorptance, which is modeled as a spectral response function utilizing Lorentzian-distributed Feibelman parameters derived from recent estimates of a 0.84% interfacial absorption. Although this model predicted distinct 1-2% reduction in reflectance centered at resonant frequencies for which the surface absorptance was modeled at, no such spectral features or localized attenuations were observed in the experimental data. The results suggest that under ambient conditions, interfacial absorption remains well below 1%, indicating that the optical signature of the photomolecular effect is either significantly weaker than previously reported or highly dependent on specific thermal and environmental conditions not present in this experiment.

1. **Introduction**

Water is a ubiquitous resource that permeates all facets of life and serves as a universal solvent. Water evaporation under natural sunlight plays a vital role in sustaining the Earth's hydrological cycle, facilitating the continuous exchange of water between land, ocean, and the atmosphere [1]. Water evaporation is equally important in industrial applications such as drying, desalination, and wastewater treatment [2-6]. Under standard conditions, bulk liquid water does not undergo a phase transition to water vapor below its boiling point, instead phase change is relegated to the surface of the water [7]. Evaporation can occur at any temperature provided there is a chemical potential gradient between the liquid's free surface and the surrounding environment [8]. The chemical potential gradient compels surface molecules to escape into the vapor phase, while vapor-phase molecules simultaneously condense back into the liquid [9]. Although both evaporation and condensation happen continuously, a net flux toward evaporation persists if the chemical potential gradient is the difference between the partial pressure of water vapor in the air and the saturation vapor pressure of water at that same temperature. Given that evaporation is intrinsically a surface phenomenon, the primary goal for enhancing this rate under solar irradiance is by localizing heat at the surface using a solar absorber (material with high absorptivity at solar wavelengths) [10, 11]. This has prompted the development of a plethora of interfacial solar evaporators and other light driven devices[12-14]. The literature is converging to the strategy of a highly porous, carbon doped three-dimensional material floating on the surface of the water, with enhancement of the capillary wicking to enable large evaporative fluxes [14-18].

Several of these materials have been claimed to beat the so-called thermal limit of evaporation [13-15, 19]:

$$\dot{m}_{Limit} = \frac{\dot{Q}_{in}}{h_{fg} + \Delta h_{sensible}} \qquad (1)$$

where $\dot{m}_{Limit}$ is the evaporation mass flux limit for a given input heat, $\dot{Q}_{in}$, and $h_{fg} + \Delta h_{sensible}$ is the combined latent plus sensible enthalpy change exhibited by the water. For 1-sun irradiance ($1 \text{ kW/m}^2$), the resulting evaporation rate is $\sim 1.5 \text{ kg/hr} \cdot \text{m}^2$ assuming no sensible heating. However, it is not fair to quantify some of the larger reported evaporative fluxes ($3 - 35 \text{ kg/hr} \cdot \text{m}^2$) against this thermal limit as three-dimensional structures have enhanced surface areas and heat fluxes. Equation (1) also does not take into account the energy transferred to the water from the environment (evaporative cooling and ambient heat transfer). The inclusion of heat transfer from the surroundings has yet to fully explain the experimentally observed "super thermal evaporation", as seen with mass and heat transfer modeling [20, 21]. Another explanation that has been offered is a reduction in the latent heat of vaporization [14, 21], but this again does not capture the enhanced evaporation rates, as shown by thermal and mass transport modeling[21].

The proposed photomolecular effect explanation by Lv et al. [20] is that in which incident light creates a steep electric field gradient at the interface, adding a significant driving force for evaporation. According to the macroscopic Maxwell equations, the normal component of the electromagnetic wave's displacement field remains continuous across the interface. The perpendicular component of the field drops by nearly half over just a few angstroms when passing from air ($\varepsilon' = 1$) to water ($\varepsilon' = 1.8$) here $\varepsilon'$ is the real part of the relative permittivity, giving rise to a pronounced electric field gradient [20]. The resulting electric field gradient interacts with the dipole moment of water clusters [22, 23], driving the cluster away from the water's surface. Experimental observations by Lv et al. [20] suggest that an interfacial optical absorptance is dependent on polarization, wavelength, and angle of incidence. Tu et al. [24] also showed that the photomolecular effect is more pronounced at lower light intensities, as water clusters are more prevalent at lower temperatures. The estimations of interfacial absorptance were based on temperature variations, and not a direct measurement of optical attenuation.

To address these gaps, the present study aims to measure the interfacial absorptance of water and its dependence on polarization of the incident light, angle of incidence (13.6° - 55.4°), and frequency (or wavelength 0.45 μm - 0.75 μm). By systematically varying these parameters, one can discern how the water–air interface responds under different electromagnetic field conditions. Such insights are not only essential for confirming and quantifying the interfacial absorption but also for understanding how, and to what extent, it contributes to phenomena like super thermal evaporation or "photomolecular" effects at an air-water interface.

## 2. Experimental Methodology

*Measurement Technique*

The approach to probe interfacial absorptance is a reference spectroscopy technique shown in Fig. 1(a) that involving adding a layer of deionized water to a reference sample and determining the difference in the polarized spectral reflectance. This is accomplished using the ratio of the polarized reflectance between a 2-mm-thick water over the substrate and that of the bare substrate, as shown in Fig. 1(b). The incidence is either transverse magentic (TM) or transverse magnetic (TE) polarized, over a range of incident angles from 13.6° - 55.4°. Specifically, the polarized reflectance of a dry substrate (silicon or platinum) is measured as a reference. A layer of water is added to the same substrate, and the polarized reflectance $R_p$ or $R_s$ is measured again. The ratio of these two reflectance signals is then compared to multilayer optical modeling. The substrates are placed in a 100-mm polystyrene petri dish on an optical table. To ensure the validity of the optical measurements, the incident beam is precisely aligned with the geometrical center of the petri dish. This captures the flattest interfacial region, effectively minimizing the curvature of the liquid surface at the container edges. To ensure complete surface wetting and a flat water layer, the substrate was initially flooded with excess deionized water that was subsequently removed to establish a thin, uniform liquid layer of 2-mm.

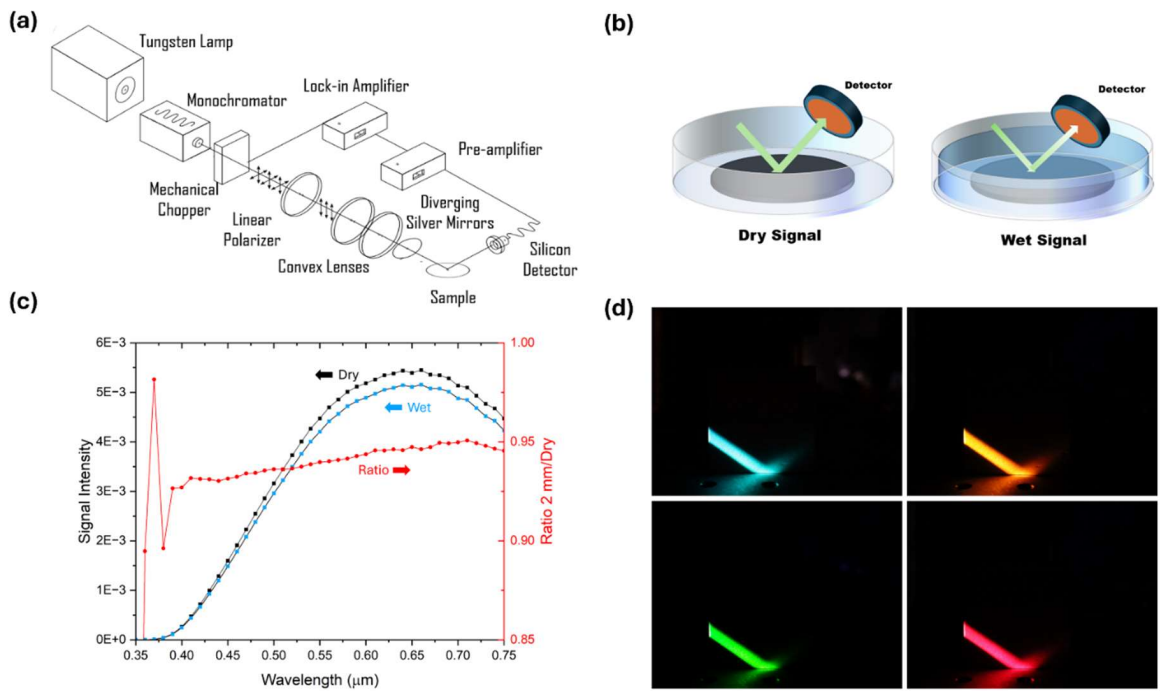

Fig. 1. (a) Measurement setup: a heated tungsten lamp is heated to serve as a broadband light source, in which individual wavelengths are spatially filtered with a monochromator to create a square tophat beam. The beam is mechanically chopped at 400 Hz before passing through a wire-grid linear polarizer. The beam is then focused on the sample through a series of convex lenses and silver mirrors. A large-area Silicon detector is placed immediately to capture the reflected light. (b) Schematic depicting the dry and wet sample placed in a plastic petri dish with beam incidence on a flat surface. (c) Reflectance signal intensity of a dry substrate and a substrate with a 2 mm water film (left y-axis), and their ratio (right y-axis) as a function of different wavelengths in the visible range. (d) Angle of incidence is measured using a highly reflective block of polytetrafluoroethylene (PTFE) to capture a photo of the incident beam, with multiple wavelengths being compared to eliminate chromatic aberration and dispersion.

The optically thick water layer induces multiple internal reflections and beam displacement, necessitating a large-area photodiode to ensure complete collection of the reflectance signal.

Consequently, an OSI Optoelectronics PIN-25D photodiode was positioned in close proximity to the sample to bypass the chromatic dispersion and beam path errors that could be introduced by subsequent collection optics. These multiple spread-out reflections, coupled with a loss of phase coherence, preclude the use of commercial ellipsometry; furthermore, such instruments are generally restricted to angles of incidence exceeding 45° rendering them unsuitable for the near-normal characterization.

As shown in Fig. 1(a), a tungsten lamp serves as a broadband light source, with 8 A current running through the filament to heat it to incandescence. The lamp is pre-heated for 2 hours to ensure thermal stability and constant resistivity. The emitted light is spatially filtered with a monochromator to isolate individual wavelengths, producing a square tophat beam profile. This ensures that a narrow band of wavelengths are incident on the sample to probe them individually as opposed to using a broadband beam and an interferometer. The narrowband beam is mechanically modulated at a frequency of 400 Hz with a chopper. The signal from the optical chopper is supplied to a lock-in amplifier to enhance the signal-to-noise ratio and isolate the response from ambient radiation. The beam is then passed through a wire-grid linear polarizer to produce TE or TM light. A series of convex lenses and protected silver mirrors then create a parallel beam incident on the sample. The reflected light is subsequently collected by the photodetector. No polarization analyzer is used after the reflectance, and cross-polarization is not considered. The signal from the photodetector is then fed into a preamplifier before going into the lock-in amplifier.

An example of the measured reflectance signal intensity is shown in Fig. 1(c) as a function of wavelength for the dry (black curve) and wet (blue curve) sample, along with their calculated ratio (red curve). Both spectra exhibit a broadband response that increases with wavelength, reaching a maximum near 0.65 µm, followed by a gradual decline toward 0.75 µm. This matches the intensity spectrum of the tungsten lamp. The wet condition shows a reduction in signal intensity relative to the dry state, which is attributed to additional reflection and absorption losses introduced by the water layer. The intensity ratio (Wet/Dry) is calculated across wavelengths, and it is this ratio that the experiment is based on. The wavelengths measured were 0.45 µm to 0.75 µm in 0.01 µm increments, and at each step 10 readings are averaged over a period of 10 seconds. Then five separate wet/dry trials were performed for each incident angle per substrate to gather the ensemble of data for a particular incident angle and polarization.

*Theoretical Model*

The reflectance ratio needs to be compared to an optical model to provide insight into the existence of interfacial absorptance. The beam angle of incidence on the air-water surface greatly impacts the theoretically predicted result. The angle of incidence is measured using a highly reflective block of polytetrafluoroethylene (PTFE) to capture a photo of the incident beam. This is processed with AutoCAD software to measure the effective angle and beam convergence. The initial optical path and lens choices were designed with a 45° angle of incidence, in mind resulting in a relatively parallel beam. When the beam is less collimated, the varying angles of incidence can be accounted for by integrating over the solid angle to determine the effective incidence angle. The angles were compared across different wavelengths to check for chromatic aberration and dispersion, as shown in Fig. 1(d).

Two substrates are utilized in this study: (i) a 75 mm boron-doped silicon wafer, and (ii) the same 75 mm silicon wafer with a 10 nm layer of titanium to serve as an adhesion layer for 400 nm of platinum. The titanium and platinum are sputtered in a low-vacuum chamber using a DC magnetron gun to strike plasma in an argon environment creating the conformal coatings on the silicon wafer. Platinum was chosen due to its high broadband reflectance and stability. Commercial mirrors typically have protective coatings, introducing thin film interference effects, complicating the optical modeling and obfuscating the phenomena of interest. A substrate like gold is stable, but it has a color and its optical properties change drastically from blue to green wavelengths of light which is of particular interest to probe the photomolecular effect. Materials like aluminum and silver do not have a color, but oxidation, sulfidation, and passivation layers complicate their theoretical modeling and obfuscate the existence of any interfacial absorptance.

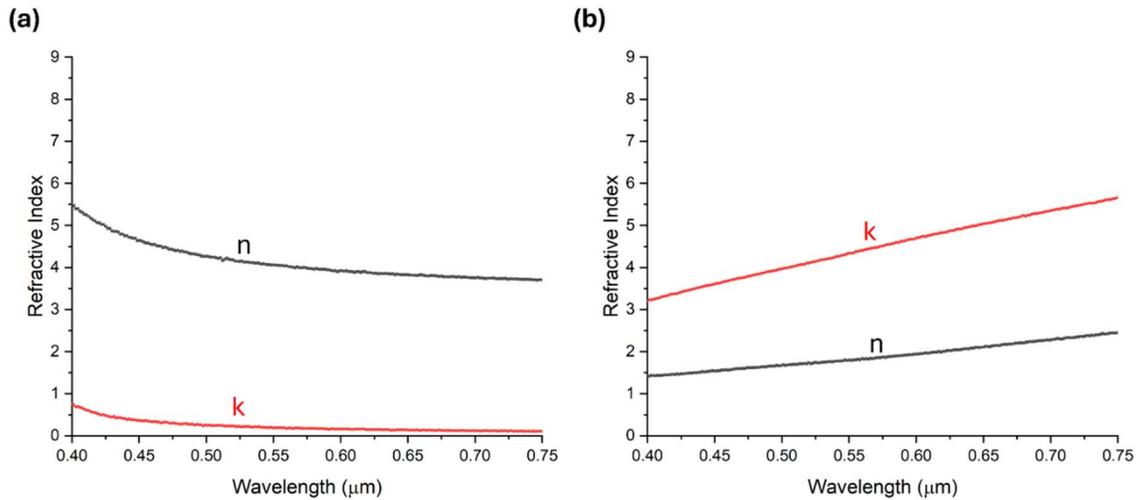

Fig. 2. Pseudo-optical properties of the a) silicon substrate and b) the platinum substrate (both in dry state). The substrates show consistent optical properties with respect to angle of incidence, spatial location, and time, with a standard deviation of 0.5%.

Pseudo-optical (*i.e.,* effective) properties of the substrates were measured using a Woollam M2000 ellipsometer. This includes the amplitude, $\Psi$, and the phase shift, $\Delta$, of the ratio of Fresnel coefficients for p- and s-polarization $r_p$ and $r_s$:

$$\frac{r_p}{r_s} = \tan(\Psi)e^{(-i\Delta)} \qquad (2)$$

Assuming a semi-infinite plane, the pseudo-optical properties of the substrate can be determined with the Fresnel reflection coefficients:

$$r_p = \frac{\tilde{n}_2^2 k_{1z} - \tilde{n}_1^2 k_{2z}}{\tilde{n}_2^2 k_{1z} + \tilde{n}_1^2 k_{2z}} \quad \text{and} \quad r_s = \frac{k_{1z} - k_{2z}}{k_{1z} + k_{2z}} \qquad (3)$$

Where $\tilde{n}_{1,2}$ is the refractive index of the incident and transmission media respectively, and $k_{1,2z}$ is the respective media's z-component of the wave vector. This accounts for surface roughness and other properties that would change the refractive index of a pure bulk material. A cartesian grid of points were measured across each substrate at different incidence angles ranging from 45° to 75°, with the average reported in Fig. 2 for both silicon and platnium. The standard deviation of the psuedo-optical properties is less than 0.5% for both substrates.

The psuedo-optical properties are then used to calculate the reflectance ratio between the dry substrate and wet substrate (with a 2-mm layer of water). The reflectance, $R$, of the water-substrate multilayer structure is:

$$R(d) = \rho_1 + \frac{\tau_1 \tau_{int}^2 \rho_{sub} \tau_2}{1 - \tau_{int}^2 \rho_{sub} \rho_2} \mid d > 0$$
$$R(d) = \rho_{sub} \mid d = 0 \qquad (4)$$

Here, $\rho_{sub}$ is the reflectivity of water-substrate interface, $\rho_1$ and $\tau_1$ are the reflectivity and transmissivity of the air-to-water interface respectively, $\rho_2$ and $\tau_2$ are the reflectivity and transmissivity of the water-to-air interface respectively, and $\tau_{int}$ is the internal transmissivity through the water layer:

$$\tau_{int} = \exp(-2\operatorname{Im}(k_{2z})d) \tag{5}$$

where $d$ is the thickness of water layer, with water's properties obtained from [25]. This is then used to model the ratio of the wet to dry reflectance:

$$\text{Ratio} = \frac{R(d=2\text{ mm})}{R(d=0\text{ mm})} \tag{6}$$

## 3. Results and Discussion

Five experimental trials were initially performed to measure the wet-to-dry signal ratio for a silicon substrate at an incidence angle of 44°. As shown in Fig. 3, the experiments are in close agreement with the theoretical thick-film model assuming zero surface absorptance (represented by the solid and dashed black curves for TM and TE polarizations, respectively). Subsequently, measurements were conducted at a larger incidence angle of 55.4°, where the photomolecular effect is reported to be more pronounced [6, 13]. At this angle, however, the silicon substrate measurements deviate significantly from theoretical values for the TM polarization as the system approaches the Brewster angles for the water and silicon interfaces.

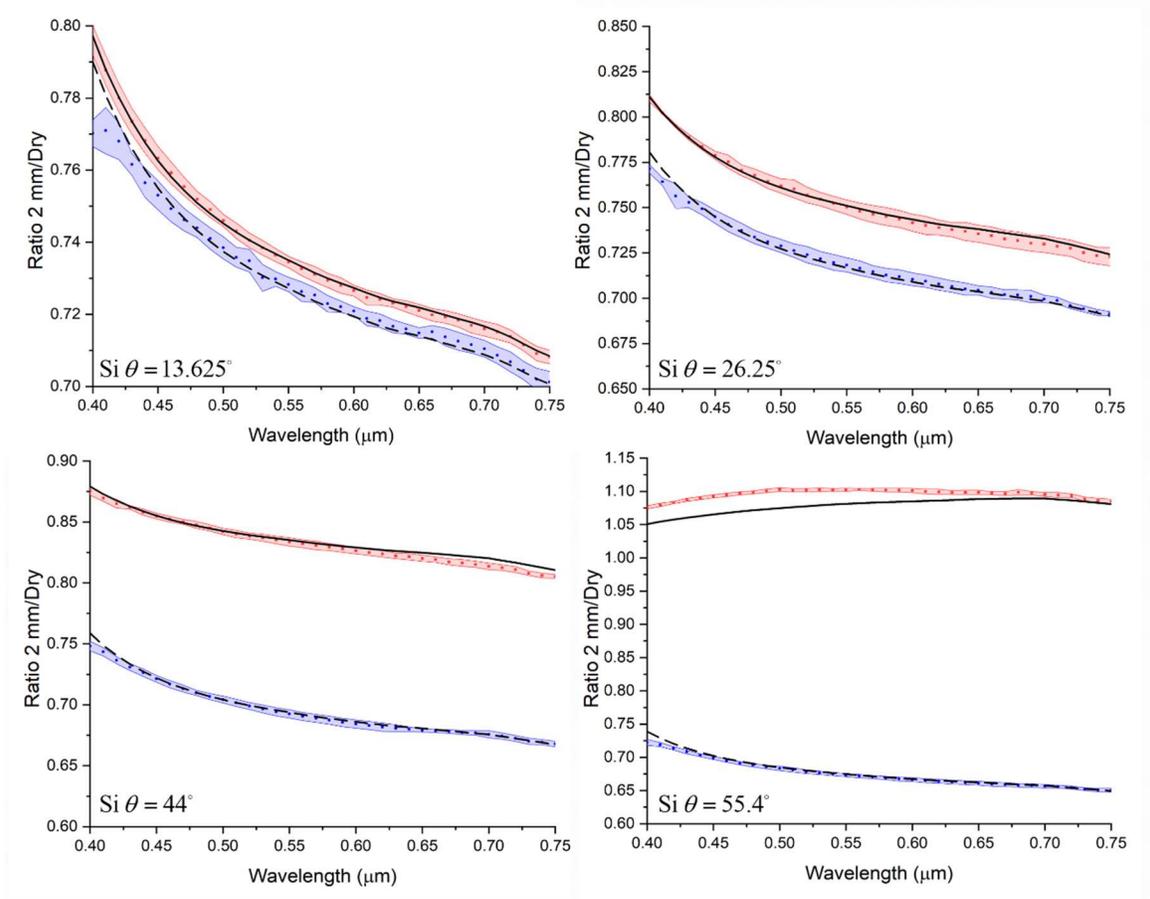

Fig. 3. Theoretical reflectance ratio (wet-to-dry signal) for a silicon substrate for TE (dashed black curve) and TM (solid black curve) incident light without surface absorptance modeled compared to experimentally measured values for TE (blue data points) and TM (red data points). The shaded bands around the measurement data represent the standard deviation across five independent trials.

The thick film multilayer reflectance model suggests that the sensitivity of the reflectance ratio to the angle of incidence increases as the angle increases, with a weak dependence on increasing wavelength. This theoretical sensitivity is depicted in Fig. 4, where it is represented as the partial derivative of the reflectance ratio with respect to the incidence angle $(\partial R/\partial \theta)$, estimated using a central finite-difference. The partial derivative increases with angle significantly for TM polarization; above incidence angle of 55°, the partial derivative of interest is greater than 0.04, a behavior that is markedly less pronounced in the (TE) state. These high-sensitivity regimes, coupled with the experimental results observed at 55.4° (Fig. 3), prompted the evaluation of alternative substrates, namely platinum. Unlike silicon, the partial derivative for platinum remains lower across the entire range of angles for both the TE and TM states. Furthermore, platinum exhibits a subtle wavelength dependence between 65° and 80°, a phenomenon primarily driven by the volumetric absorption of the aqueous medium. Notably, a localized region of minimum sensitivity exists for platinum near 82°, where the partial derivative undergoes a sign reversal.

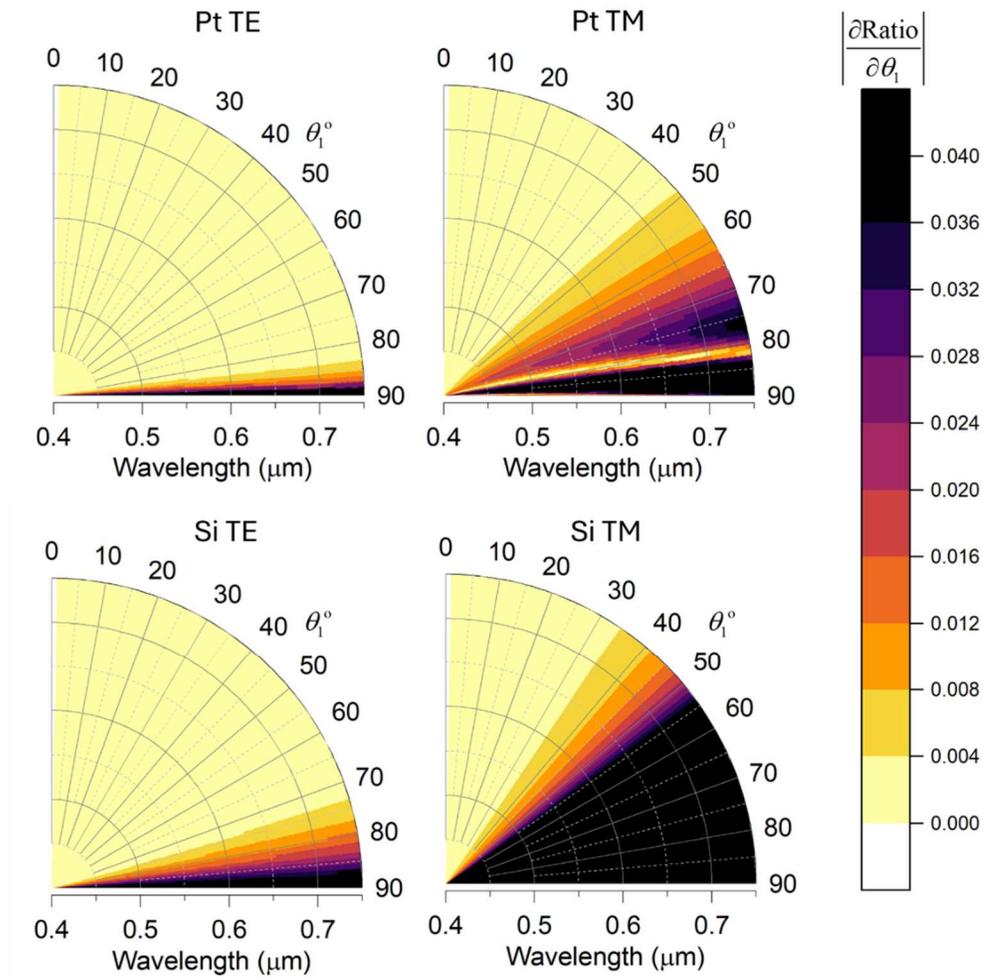

Fig. 4. Sensitivity of the theoretical reflectance ratio to incident angle changes for TE and TM polarizations with platinum and silicon substrates.

The heightened angular sensitivity and subsequent challenges in maintaining precise optical alignment beyond 55° confined reliable data collection to smaller incident angles for both substrates. Additional measurements were obtained for the silicon substrate at 13.625° and 26.25° as shown in Fig. 3. For comparative analysis, data were also collected for the platinum substrate at these same incidence angles of 13.625°, 26.25°, 45°, and 55.4°, as illustrated in Fig. 5. For the silicon substrate, the experiments are in strong agreement with the multilayer model at lower angles, specifically 13.625°, 26.25°, and 44°. However, a notable divergence occurs at 55.4°, attributed to the combined effects of diminished reflectance, and the steep increase in angular sensitivity previously discussed (Fig. 4). These factors exacerbate the impact of beam alignment uncertainties, leading to the observed discrepancy between the theoretical model and the experimental results. Consistent across all measurements, and most distinguishable at 13.625°, is the observation of the signal drop at wavelengths between 700-750 nm due to the volumetric absorption of water, demonstrating the system's capability to resolve absorptance through a ~4 mm effective path length in water.

The platinum measurements show agreement within the standard deviation of the trials at 26.25°, 45°, and 55.4°. However, a distinct trend was observed for the TM polarization case at 45° and 55.4°, where the measured reflectance values exhibited a small but consistently lower broadband shift than the theoretical predictions. These shifts can be attributed to small deviations in the experimental path length. In contrast, the 13.625° platinum measurements showed a more pronounced discrepancy; for both TM and TE polarization states, the theoretical values fell outside the standard deviation of the experimental data.

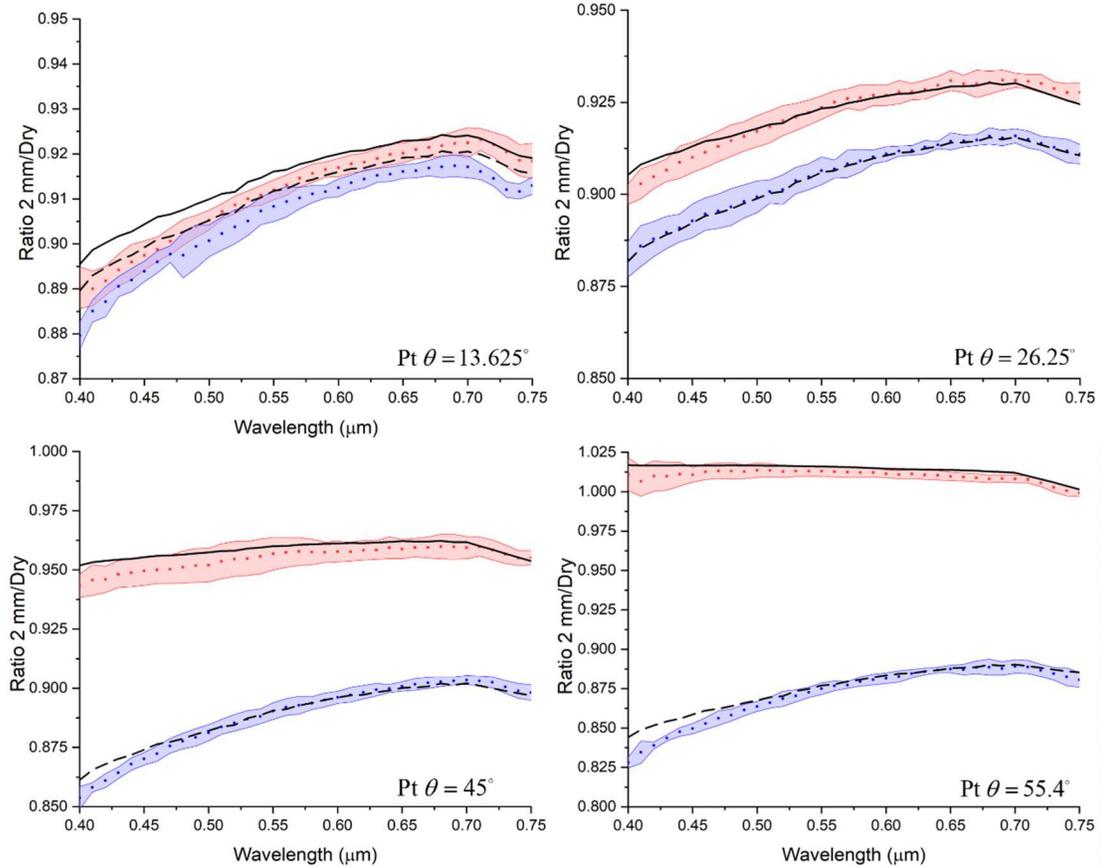

Fig. 5. Theoretical reflectance ratio (wet-to-dry signal) for a platinum substrate for TE (dashed black curve) and TM (solid black curve) incident light without surface absorptance modeled compared to experimentally measured values for TE (blue data points) and TM (red data points). The shaded bands around the measurement data represent the standard deviation across five independent measurement trials.

At these lower angles of incidence, the distinction between the TM and TE polarization states diminishes significantly. Consequently, the polarization-dependent Fresnel coefficients converge at normal incidence, causing the signal ratios to become nearly degenerate. The convergence of these states explains why the experimental values for both polarizations are tightly clustered at smaller angles. This observation highlights a fundamental duality in the proposed measurement technique: while larger incidence angles have a higher distinguishability of polarization, they simultaneously have increased susceptibility to optical deviations and alignment uncertainties. Conversely, smaller incidence angles offer greater robustness against path length alignment uncertainty, albeit at the cost of a reduction in the distinctness of the polarization states.

Although the experimental data align closely with the theoretical model assuming zero surface absorption, it is essential to quantify the potential impact of any such absorption on the system's response. Generally, two primary methods are employed to model the surface absorption of media. The traditional approach involves the addition of a discrete, absorbing layer at the interface, as characterized by the classic McIntyre and Aspnes model [26]. Another method is introducing a spectral response function called Feibelman parameters in the standard macroscopic Maxwell equations, with a modification of boundary conditions to serve as self-consistent field discontinuities [27]. These Feibelman parameters model nanoscale interfacial optical phenomena, and Chen used this approach to model the photomolecular effect and the impact on the reflection and transmission coefficients [28, 29].

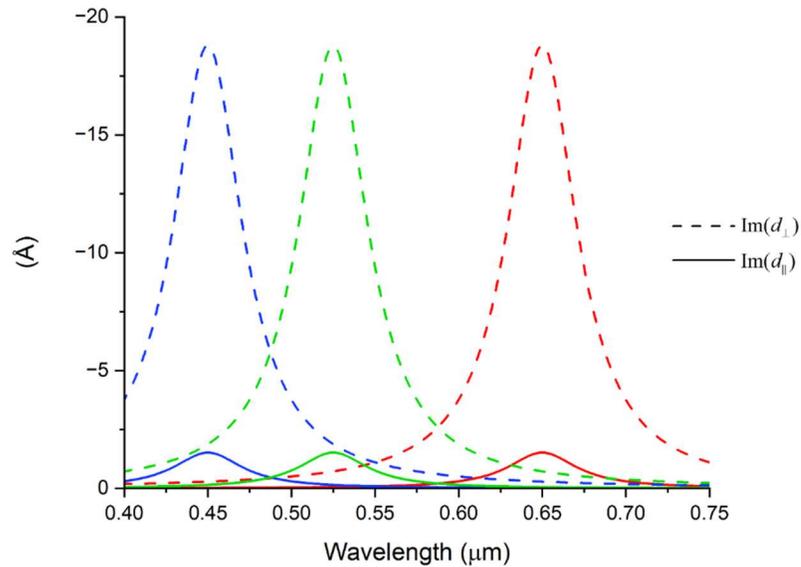

Fig. 6. Magnitude of imaginary parts of Feibelman parameters when the permittivity of the substrate is larger than the permittivity of the incident medium; the imaginary parts of the Feibelman parameters must be negative.

The specific values reported for the Feibelman parameters depend directly on the magnitude of the interfacial absorptance, which were derived based on temperature measurements and convective heat transfer modeling yielding an inferred absorption of 0.84% for a wavelength of 532 nm [20, 28]. Then applying these same absorption values at 520 nm, the imaginary values were reported to be $d_\perp \approx 18.8$ Å and $d_\parallel \approx 1.52$ Å [28]. The imaginary Feibelman parameters, representing system dissipation, are the only components discernible from absorption measurements. Determining the full complex response without phase data requires the use of Kramers-Kronig relations [30]. The exact mechanism of the dissipation in water associated with a photomolecular effect is still being investigated [31]. Li et al. subsequently extended this methodology from a single interface to a practical air-water-substrate configuration, illustrating the interplay between surface, bulk water, and substrate absorptions[32].

The proposed values of the Feibelman parameters are used to modify the thick film optical model. Instead of utilizing a constant Feibelman parameter across all wavelengths as previously investigated, a spectral Lorentzian distribution is used to capture the reported frequency dependent behavior of the photomolecular effect and subsequent surface absorption [33]. The utilization of a Lorentzian distribution is dependent on the assumption of homogenous broadening of the photomolecular effect optical signature, which suggests a quasi-particle or molecular cluster excitation mechanism at the liquid-vapor interface. Unlike a Gaussian profile, which typically represents inhomogeneous broadening, the Lorentzian line shape effectively captures damped driven-oscillator behavior of electromagnetic transitions. This allows the model to account for the 'tails' of the absorption peak, ensuring that off-resonant spectral contributions are not abruptly truncated and are still included in the multilayer model. Again, this modeling approach aims to capture the wavelength dependent behavior reported in previous studies [20].

The resulting distributions for $d_\perp$ and $d_\parallel$ are depicted in Fig. 6 at three different central wavelengths (colors). The resonant frequency for the photomolecular effect was reported to be associated with green light, as this wavelength showed the largest temperature increase attributed to a surface absorptance [20]. In contrast, the blue (450 nm) and red (650 nm) wavelength models are included solely as comparative baselines to assess the theoretical impact of the measurement ratio dependency on wavelength. These secondary wavelengths serve as a control, demonstrating how the system response would deviate from the resonant green-light behavior, despite the current lack of experimental evidence for the effect at these specific wavelengths.

The influence of Feibelman parameters on the multilayer model is illustrated in Figs. 7 and 8 for silicon and platinum substrates respectively. As similarly shown in [28], across all incident angles, the impact of the chosen Feibelman parameters on silicon's reflectance ratio is minimal for TE polarization, resulting in an almost negligible drop in the reflectance ratio. Conversely, platinum exhibits a more pronounced drop in TE reflectance. This is influenced by the high reflectance of the metallic interface and the subsequent attenuation as light propagates from water to air. This drop diminishes at larger incidence angles and longer wavelengths, consistently following the selected Lorentzian distribution. However, no such features are observed in any of the experimental data.

The TM reflectance ratio for both silicon and platinum drops significantly for all three colors of surface absorptance and as expected increases with larger incident angles. The Lorentzian Feibelman distributions show a 1-2% drop for 45° and 55.4° for both silicon and platinum with platinum having a more pronounced effect much larger than the standard deviation of the measurement data. The experimental curves remain smooth and monotonic across the visible spectrum, lacking the resonant absorption peaks that the Feibelman parameters would theoretically induce. This qualitative discrepancy suggests that the magnitude of the $d_\perp$ and $d_\parallel$ components derived from Chen 2024 may overstate the actual interfacial dissipation present in this specific experimental configuration [28].

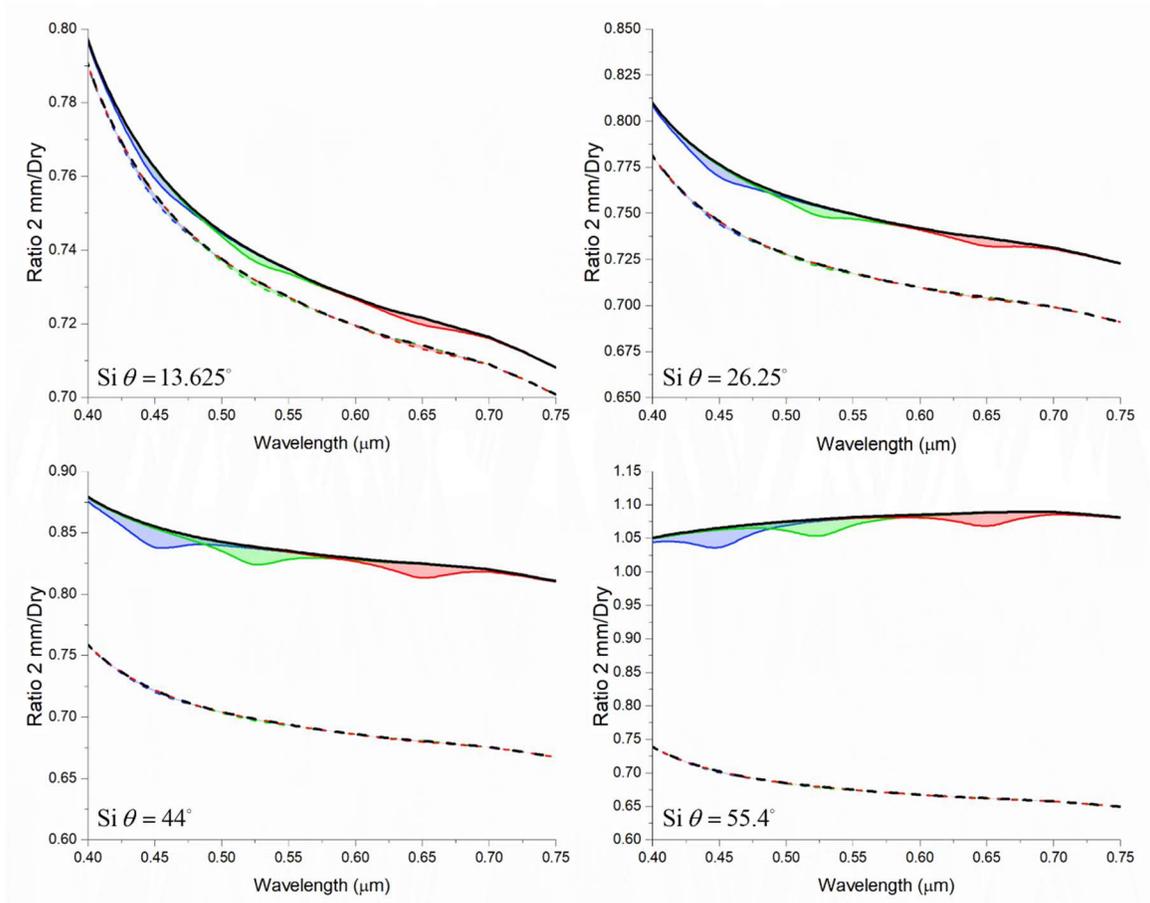

Fig. 7. Theoretical reflectance ratio (wet-to-dry signal) for a silicon substrate for TE (dashed black curve) and TM (solid black curve) incident light without surface absorptance modeled compared to surface absorptance models (blue, green and red curve; solid denoting TM and dashed denoting TE). The shaded regions of the respective colors show the deviation from the black line influenced by the Feibelman parameters shown in Fig. 6.

Furthermore, these experiments were conducted under ambient conditions without specific controls for water temperature or localized humidity. Notably, there were no temporal controls implemented to account for the theorized mechanism of molecular clusters evaporating from the water surface under illumination and recondensing under dark conditions. The lack of observed attenuation in the measurement data leads to the conclusion that no measurable interfacial absorption layer ~1% is present under our experimental conditions and no frequency dependent behavior was observed. This suggests that the photomolecular effect may either require more specific conditions manifest, or that its optical signature is too faint to perturb the Fresnel response of a standard water-vapor interface.

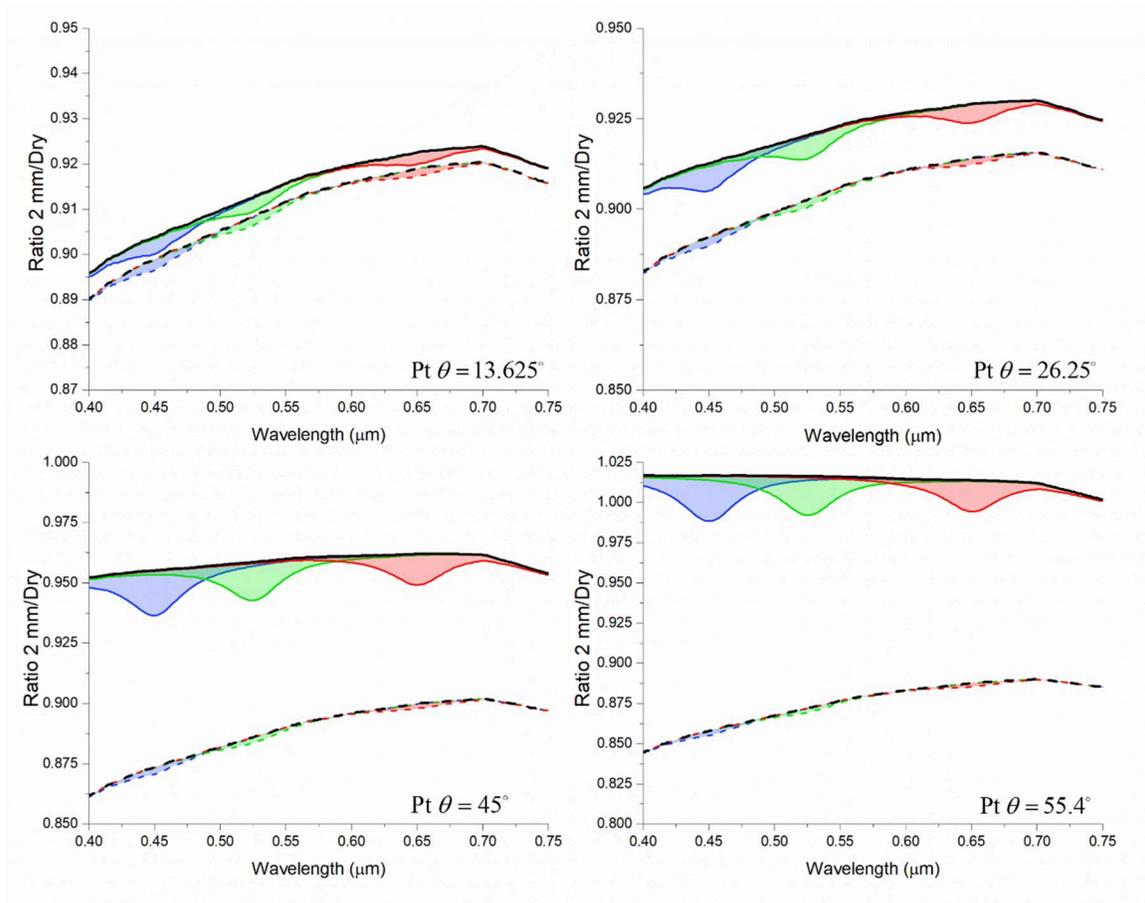

Fig. 8. Theoretical reflectance ratio (wet-to-dry signal) for a platinum substrate for TE (dashed black curve) and TM (solid black curve) incident light without surface absorptance modeled compared to surface absorptance models (blue, green and red curve; solid denoting TM and dashed denoting TE). The shaded regions of the respective colors show the deviation from the black line influenced by the Feibelman parameters shown in Fig. 6.

## 4. Conclusion

In summary, this research evaluated the optical response of the water-air interface to determine if resonant surface absorption, attributed to the photomolecular effect, could be resolved through polarization-dependent reflectance measurements. By comparing silicon and platinum substrates, we identified that while larger incidence angles enhance reflectance signal sensitivity, they simultaneously amplify susceptibility to alignment uncertainties, whereas smaller angles provide robustness at the cost of polarization distinctness. The experimental measurements for both substrates remained remarkably consistent with classical Fresnel theory across the visible spectrum. Specifically, the high-sensitivity TM polarization state failed to exhibit the absorption "dips" predicted by previously reported Feibelman parameters.

The absence of these spectral features leads to the conclusion that no measurable interfacial absorption layer ~1% was present under our experimental conditions. These findings suggest that the photomolecular effect may be a conditional phenomenon requiring precisely controlled thermal gradients or localized humidity levels to manifest a detectable optical signature. Future work should focus on synchronized thermal-optical monitoring to further define the limits of this interfacial dissipation and reconcile the discrepancy between convective heat transfer models and direct electromagnetic observations.

**CRediT authorship contribution**

**Preston Bohm:** Conceptualization, Investigation, Methodology, Data curation, Visualization, Writing- Original Draft, Writing–review & editing; **Mingjun Li:** Investigation, Methodology, Data curation, Validation, Visualization, Writing–review & editing; **Akanksha K. Menon:** Resources, Funding acquisition, Supervision, Writing–review & editing; **Zhuomin M.**

**Zhang:** Conceptualization, Methodology, Supervision, Funding acquisition, Writing–review & editing.

**Declaration of competing interests**

The authors declare that they have no known competing financial interests or personal relationships that could have appeared to influence the work reported in this paper.

**Data availability**

Data will be made available upon request.

**Acknowledgments**

This work was made possible by the Woodruff Launch program and the J. Erskine Love Jr. Endowed Professorship from the Woodruff School of Mechanical Engineering at Georgia Tech.